# Strong coupling between magnons confined in a single magnonic cavity

Changting Dai, Kaile Xie, Zizhao Pan, and Fusheng Ma[*]

Jiangsu Key Laboratory of Opto-Electronic Technology, Center for Quantum Transport and Thermal Energy Science, School of Physics and Technology, Nanjing Normal University, Nanjing 210023, China

*Corresponding author. E-mail: phymafs@njnu.edu.cn

## Abstract

Strong coupling between magnons and cavity photons was studied extensively for quantum electrodynamics in the past few years. Recently, the strong magnon-magnon coupling between adjacent layers in magnetic multilayers has been reported. However, the strongly coupled magnons confined in a single nanomagnet remains to be revealed. Here, we report the interaction between different magnon modes in a single magnonic cavity. The intermodel coupling between edge and center magnon modes in the strong coupling regime was approached with a maximum coupling strength of 0.494 GHz and cooperativity of 60.1 with a damping of $1 \times 10^{-3}$. Furthermore, it is found that the coupling strength is highly dependent on the geometric parameters of the magnonic cavity. Our findings could greatly enrich the still evolving field of quantum magnonics.

# Introduction

Light-matter interaction in the strong coupling regime has been widely investigated for quantum information and quantum communication. Especially, the interaction between cavity photons and phonons, excitons, magnons, plasmons, as well as superconducting qubits has been experimentally demonstrated in the strong coupling regime.[1–6] Since the theoretical prediction by Ö. O. Soykal and M. E. Flatté in 2010,[7,8] the strong coupling between cavity microwave photons and magnons in yttrium iron garnet has been demonstrated in various experimental systems.[6,9–13] The microwave cavity usually has the dimension on the order of tens millimeter. Besides, the coupling strength between microwave photons and magnons is proportional to the square of the number of spins, that is, $g \propto \sqrt{N}$.[9,14,15] In order to increase the coupling strength, the number of spins in the magnetic material is usually required to be large enough, i.e., the volume of the magnetic material usually in millimeters. Hence, the size of microwave cavity and magnet in the strong magnon-photon coupling is restrictive in device miniaturization and CMOS-integration.

In view of the above shortcomings, it has become a hot topic to search for a nanometer resonator to replace the microwave cavity. The magnon modes confined in magnetic nanostructures show the potential to provide cavity modes. Magnons are the quanta of spin waves, which are the collective excitations of spins in magnetic materials. Magnon has the potential to implement high energy-efficient and low heat dissipation devices on the micron to nanometer scale, which due to its charge-less diffusion, long coherent distance and time, and the available information processing frequency can reach gigahertz and above in magnetic materials.[16] In the last decade, the extensive study of encoding information using both confined and propagating magnons arises an emerging field in spintronics - magnonics.[10,17–22] Inspired by the exciting achievements of quantum optics, an open question is whether magnons can be used to perform information processing and storage on the quantum level. Very recently, the strong interlayer magnon-magnon coupling in spatially separated metal-insulator hybrid multilayers has been experimentally demonstrated.[15,22-24]

Exchange enhanced direct magnon-magnon coupling has been found in a single crystal material at ultra-low temperature.[25-28]

These exciting achievements has gradually developed into a new discipline of quantum magnonics.[29–31] In which the magnons interact coherently with the elementary excitations of matter to obtain the quantum phenomena of magnonics. An open question is that whether there is a possibility to realize the strong coupling between magnons confined in a single nanomagnet since the magnons in finite nanomagnets are copious, i.e. geometry confined volume modes and localized edge modes.[32–36] Although the evolution of various magnon modes in finite nanomagnets has been widely studied, i.e. rectangular,[37-39] square,[40,41] and triangular[42] micro/nanomagnets, the investigations of strong coupling between these modes are still rare. Only recently, the strong magnon-magnon coupling in $Ni_{80}Fe_{20}$ nanocross array has been experimentally studied by ferromagnetic resonance.[43]

In this work, we numerically investigated the interactions between various magnon modes confined in a magnonic cavity in the form of an irregular hexagonal dot (IHD). Interestingly, it is found that the character of strong coupling, anticrossing, appears in the frequency-field (*f-H*) dispersions. For damping $\alpha < 7.65 \times 10^{-3}$, the magnon-magnon interaction can approach the strong coupling regime. The observed intermodel coupling is attributed to the interaction between edge and center magnon modes. A maximum coupling strength of 0.494 GHz and a cooperativity of 60.1 with a damping of $1 \times 10^{-3}$ can be achieved. Furthermore, the coupling strength was found to be highly dependent on the geometric parameters of the magnonic cavity. Our findings could provide a new magnonic platform for exchanging quantum information between strongly coupled magnons.

**Model and Calculation**

Two types of magnonic cavities, confining magnons as an analog to microwave cavity, were modeled as illustrated in Fig. 1(a) and Fig. 1(b), respectively. Figure 1(a) presents the regular rectangular dot (RRD) with the dimension of length *L*, width *W* and thickness *d*. While, Figure 1(b) shows the IHD with length *l*, width *W* and thickness *d*. The sharp ends of IHD are in the form of isosceles triangles with a top angle *θ* as shown in Fig. 1(b). The calculation of magnon spectra was performed by micromagnetic simulations using MuMax3.[44] The magnonic cavities are discretized into cubic cells with a size of 5 nm × 5 nm × 5 nm (along *x*-, *y*- and *z*-axis). During the simulations, the thicknesses of the cavities are fixed at 5 nm, while the width and length are varied. The magnetic material parameters used are these of CoFeB[45,46] that has low magnetic damping as follows: saturation magnetization $M_s = 1.2 \times 10^6$ A/m, exchange stiffness $A = 1.1 \times 10^{-11}$ J/m, and uniaxial anisotropy constant $K_u = 5$ kJ/m$^3$ along the width of the magnonic cavities. The damping constant $\alpha = 1 \times 10^{-3}$ unless otherwise noted. To calculate the magnon spectra, a two-step simulation was performed for each external field $H_{ext}$. Firstly, a static simulation was carried out to get the ground state of the magnetization determined by minimizing the total energy of the simulated magnetic volume. Secondly, a dynamic simulation starting from the ground magnetization state was done with a radio frequency perturbation field $h_{rf}$ applied perpendicularly to the external bias field. In these two geometries, the slant spins of magnon can be obtained as schematically presented by Fig. 1(c). The $h_{rf}$ was adapted in the form of a "Sinc" function,[47-49] $h_{rf}(t) = h_0 \sin(2\pi ft)/(2\pi ft)$, where the amplitude $h_0 = 5$ mT and the cut-off frequency $f = 50$ GHz. The magnons with frequencies ranging from 0 to 50 GHz can be effectively excited. During the dynamic simulation of 10 ns, the spatially averaged magnetization $m(t)$ was saved with a time interval of 10 ps. Then, the magnon spectra at the specific field can be obtained by performing Fourier transform of the recorded $m(t)$.

**Results**

We start by considering a ferromagnetic rectangular nanodot as illustrated in Fig. 1(a), the dimension of the nanodot is $L = 560$ nm, $W = 240$ nm, and $d = 5$ nm. Figure 1(d) shows the color-plotted frequency dependence of magnons on external bias field $H_{ext}$. There are three obvious magnon modes as indicated by ①, ②, and ③. A typical magnon spectrum is shown in the left panel of Fig. 1(f) for $H_{ext}$ = 144 mT. The mode ① has the lowest frequency 13.47 GHz and the largest intensity. While, the mode ③ has the largest frequency 16.96 GHz and the lowest intensity. As shown in the right panel of Fig. 1(f), the mode profiles indicate that the observed three magnon modes are corresponding to the length confined mode with the antinode number $n$ = 0, 2, and 4 for modes ①, ②, and ③, respectively. With the bias field increasing, the frequencies of these three modes increase linearly with the relative intensity unchanged. The results shown in Fig. 1(d) is similar to that as reported for ferromagnetic nanodots.[35]

Next, we considered a structure of IHD as shown in Fig. 1(b), whose dimension is of $l$ = 320 nm, $W$ = 240 nm, $d$ = 5 nm and $\theta$ = 90°. Figure 1(e) shows the color-plotted frequency dependence of magnons on external bias field $H_{ext}$. A typical spectrum with three magnon modes is shown in the left panel of Fig. 1(g) for $H_{ext}$ = 144 mT. In contrast to the *f-H* dispersion of RRD in Fig. 1(d), the hallmark of strong coupling, anticrossing, was observed clearly from the dispersions of modes ① and ② as shown in Fig. 1(e), i.e. the intensity of mode ①/② decreases/increases with external field increases. For $H_{ext}$ = 144 mT, the intensities of modes ① and ② are similar as shown in the left panel of Fig. 1(g). The mode profiles of the three magnon modes are shown in the right panel of Fig. 1(g) with the frequencies of 12.97 GHz, 13.57 GHz and 15.36 GHz, respectively. Hereafter, we will only consider the modes ① and ② since the mode ③ is much weaker and does not interact with modes ① and ②. For mode ①, the dynamical magnetization is strongly localized in the left and right narrow edges of the cavity. Therefore, mode ① is the edge magnon (EM) mode since it is trapped in the physical edges by the nonuniform internal field. While for mode ②, the dynamic magnetization mainly distributes in the central area of the cavity, so it can be labeled as center magnon (CM) mode. Therefore, the anticrossing or modes repulsion in Fig. 1(e) is attributed to the interaction between EM and CM modes confined in the IHD cavity. The hybridization of magnon modes occurs when the frequencies of these two modes are close to each other. The dominant inducement of the anticrossing phenomenon is due to the dipole-dipole interactions.[15,50]

To illustrate the strong coupling phenomena concretely, we present the representative magnon spectra at various fields as shown in Fig. 2(a). The lower frequency EM mode has higher intensity than that of the higher frequency CM mode at $H_{ext}$ = 120 mT. By increasing $H_{ext}$ from 120 to 170 mT, it is found that the intensity of EM/CM mode decreases/increases, respectively. For $H_{ext}$ = 170 mT, the intensity of the EM mode becomes lower than that of the CM mode. Similar characteristics were found in a single square nanomagnet by P. S. Keatley et al.[51] and in a Py microstrip by Lihui Bai et al.,[52] respectively. The turning point for the relative intensity between EM and CM modes happens at $H_{ext}$ = 144 mT where these two modes have similar intensity and we defined this bias field as the coupling field $H_g$. At the meantime, the frequency difference between the EM and CM modes has the smallest value. To certify the interaction between the EM and CM modes, we define the coupling strength $g$ as the half of the modes splitting at $H_g$, and the corresponding dissipation rates $k_{EM}$ and $k_{CM}$ as the half width at half maximum of the line broadenings of the EM and CM modes, respectively. As shown in Fig. 2(b), we obtain $g = |f_{EM} - f_{CM}|/2$ = 0.310 GHz, and the dissipation rates of EM and CM modes are $k_{EM}$ = 0.044 GHz and $k_{CM}$ = 0.043 GHz. Since $g > k_{EM}, k_{CM}$, the interaction between EM and CM modes approaches the strong coupling regime of cavity quantum electrodynamics.[9] The extent of coupling can also be denoted utilizing a unitless parameter, cooperativity, which is defined as

$$C = g^2/(k_{EM} \times k_{CM}). \tag{1}$$

Then, for the coupling between EM and CM modes as shown in Fig. 2(b), the cooperativity is $C$ = 50.8. The distinct anticrossing was shown in Fig. 2(c), which displays the frequencies of two coupled magnon modes at selected fields. This observed modes repulsion can be analog to the strong coupling between magnon and cavity photon.[6-13] The intensity variations of the EM and CM modes are presented as a function of the magnon frequencies as shown in Fig. 2(d).

Since the judgement of the coupling between the EM and CM modes are based on the relative values of $g$, $k_{EM}$ and $k_{CM}$, we plotted these three values as a function of the Gilbert damping parameter $\alpha$ changing from $13.5 \times 10^{-3}$ to $1 \times 10^{-4}$ as shown in Fig. 3(a). The dissipation rates $k_{EM}$, $k_{CM}$ are highly dependent on $\alpha$, while the coupling strength $g$ is independent on $\alpha$. To satisfy the condition of strong coupling, $g > k_{EM}$, $k_{CM}$, the damping $\alpha$ should be smaller than $7.65 \times 10^{-3}$ as shown in Fig. 3(b). Otherwise, the interaction between EM and CM modes is in the weak coupling regime.[9] Figure 3(c) visually displays the effects of Gilbert damping parameter on the coupling strength and the linewidths of coupled magnon modes at coupling field $H_g = 144$ mT. The coupling strength is independent on the damping, but the linewidths of the magnons are highly dependent on the damping values. These three damping values correspond to the three symbols in Fig. 3(d). The lowest value of $\alpha$ in CoFeB is ~0.005 as reported by A. Okada et al.,[46] this damping value was shown in the orange diamond symbol in Fig. 3(d). For the epitaxial $Co_{25}Fe_{75}$ films, the damping $\alpha$ even can go down to $1.4 \times 10^{-3}$,[53] the yellow diamond symbol in Fig. 3(d) shows the damping close to this value. In magnetic insulators, the dissipation rates of magnon modes are usually much lower, resulting in the calculated cooperativity is higher, as displayed in Fig. 3(d). The green diamond symbol represents the minimum damping $\alpha = 1 \times 10^{-4}$ in simulation, corresponding to a maximum cooperativity of 199.9. Actually, in some insulating ferromagnets, the Gilbert damping constant $\alpha$ can reach $10^{-4}$ to $10^{-5}$.[54] Therefore, it is possible to experimentally observe our reported strong coupling between magnons confined in a single finite magnonic cavity.

So far, we have studied the strong coupling between magnons in IHD with fixed dimension. To see how the geometrical parameters of the IHD affect the coupling between confined magnons, we focused on the variation of $l$, $W$, and $\theta$, respectively. Figure 4(a) presents the coupling strength $g$ and coupling center frequency $f_g = (f_{EM} + f_{CM})/2$ as a function of $l$ with $\theta = 90°$ and $W = 300$ nm. We only observed the coupling happened for $l > 110$ nm, and the coupling strength $g$ decreases from 0.494 to 0.156 GHz with $l$ increases from 110 to 630 nm. The $f_g$ oscillates with the variation of $l$. For $l = 0$ nm, there is no coupling happened as shown in Fig. 4(b). For $l = 110$ nm, the coupling strength $g$ has the maximum of 0.494 GHz, and the dissipation rates $k_{EM} = 0.058$ GHz, and $k_{CM} = 0.070$ GHz at a coupling field $H_g = 124$ mT, which accesses to the strong coupling regime and the magnon-magnon cooperativity of $C = 60.1$. For larger $l$, the coupling strength becomes smaller as shown in Fig. 4(c) and Fig. 4(d) for $l$ = 170 and 630 nm. Therefore, the variation of $l$ can highly affect the strength of the coupling between confined magnons. However, we found that the coupling strength $g$ is insensitive to the width $W$ of the IHD, which is not shown here. Lastly, we studied the effect of $\theta$ on the interaction between magnons confined in the IHD cavity with $l$ = 320 nm and $W$ = 240 nm. Figure 5 shows the color-plots of $f$-$H$ dispersions of magnons in IHDs with $\theta$ = 60°, 90°, and 120°, respectively. It is a distinctly different $f$-$H$ dispersion for IHDs with different $\theta$. We found that not only the types of magnon modes change, but also the interaction between them. The anticrossing character observed for $\theta$ = 90° does not appear clearly for $\theta$ = 60° and 120°.

## Conclusion

In summary, the interaction between magnons confined in a sole magnonic cavity has been realized in the strong coupling regime. The observed intermodel coupling is attributed to the central volume magnon mode and the edge localized magnon mode in an irregular hexagonal dot. It is found that the coupling strength is sensitive to the length and the sharp-ends' angle of the magnonic cavity, while insensitive to the width. The coupling strength can be as high as 0.494 GHz with a cooperativity of 60.1. Our findings provide a magnonic platform for investing the matter-matter strong coupling in cavity quantum electrodynamics utilizing magnons.

## ACKNOWLEDGMENTS

This work was supported by the National Natural Science Foundation of China (Grant No. 11704191), the Natural Science Foundation of Jiangsu Province of China (Grant No. BK20171026), the Jiangsu Specially-Appointed Professor, and the Six-Talent Peaks Project in Jiangsu Province of China (Grant No. XYDXX-038).

# Figures

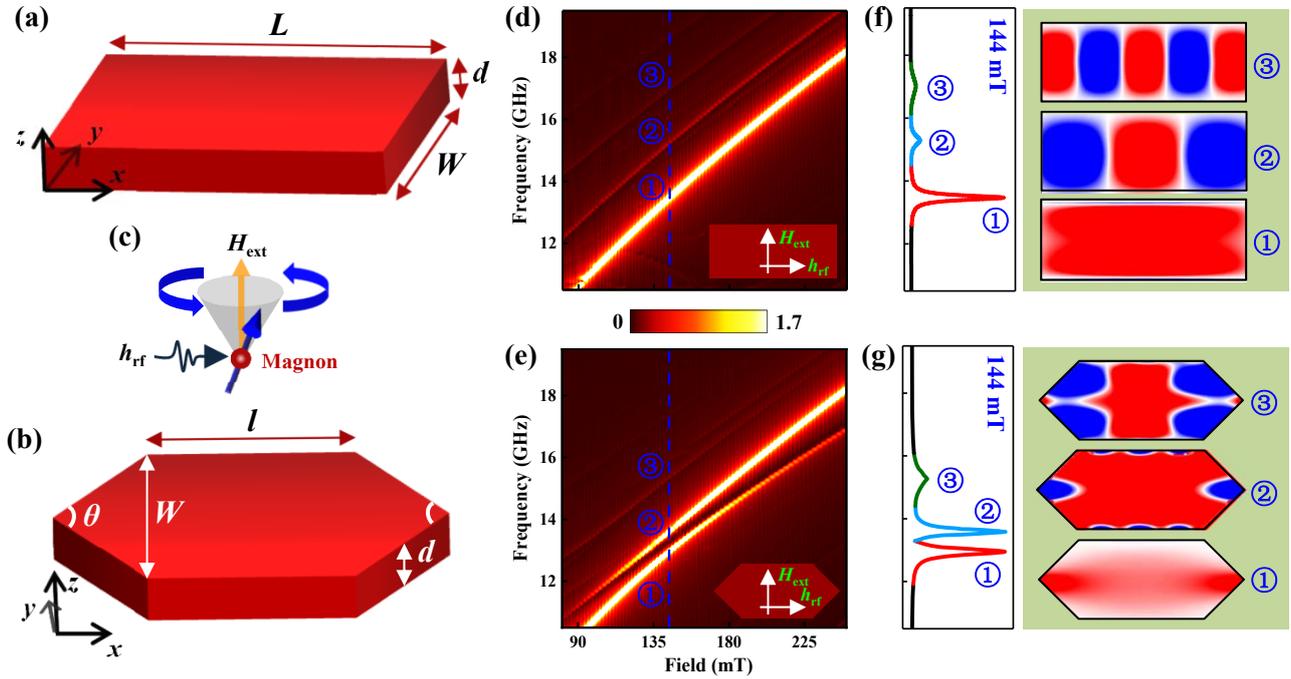

**Figure 1.** Schematic of (a) regular rectangular dot (RRD) and (b) irregular hexagonal dot (IHD). (c) Schematic of magnon excitation under a bias field $H_{ext}$ and a perturbation field $h_{rf}$. Color plots of frequency-field (*f-H*) dispersion of magnons in (d) RRD (*L* = 560 nm, *W* = 240 nm) and (e) IHD (*l* = 320 nm, *W* = 240 nm, *θ* = 90°), respectively. Insets indicate the direction of bias and perturbation fields. Typical magnon spectra of (f) RRD and (g) IHD at selected field labelled as vertical dashed lines in (d) and (e) and the corresponding mode profiles.

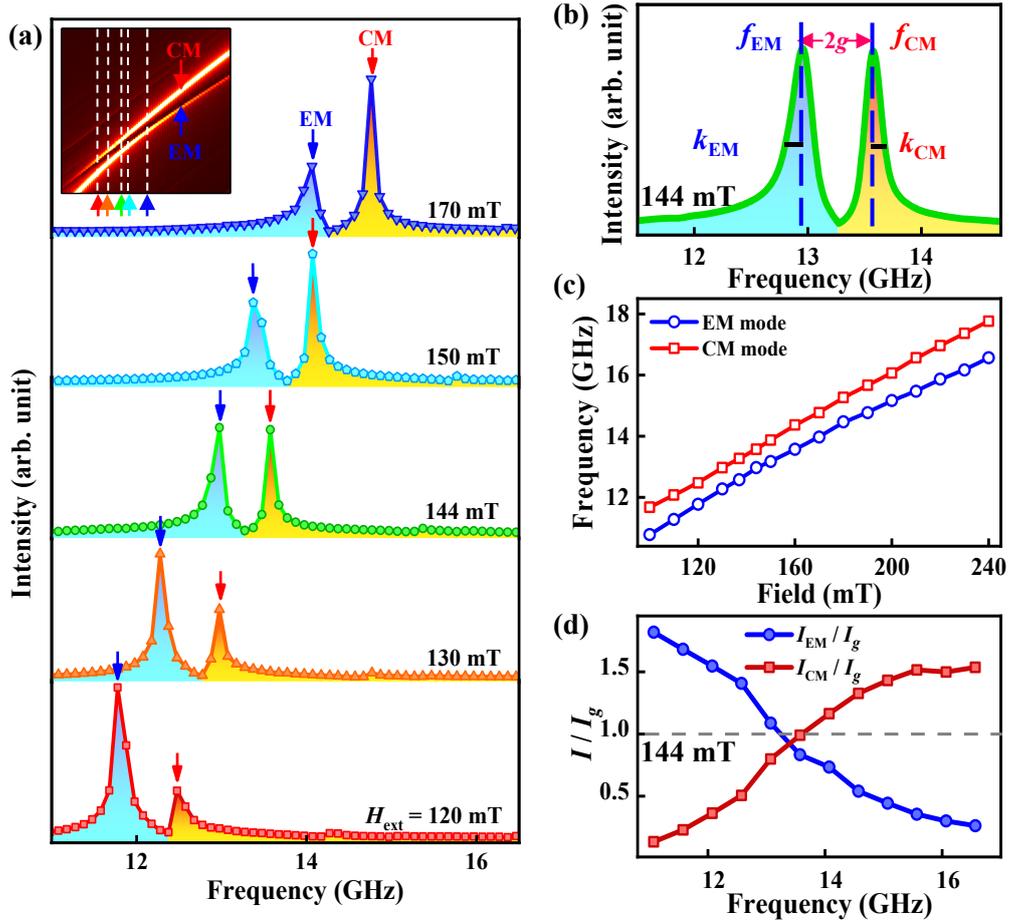

**Figure 2.** (a) The spectra of confined magnon modes in IHD ($l$ = 320 nm, $W$ = 240 nm, $\theta$ = 90°) under various bias field $H_{ext}$. The inset is the corresponding color plot of spectra as shown in Figure 1e. (b) Magnon spectrum at the coupling magnetic field $H_g$ = 144 mT, where the two magnon modes have similar intensity. The coupling strength $g$ is defined as half of the modes splitting, and the corresponding dissipation rates $k_{EM}$ and $k_{CM}$ are defined as the half width at half maximum of the line broadenings of edge magnon (EM) and center magnon (CM) modes, respectively. (c) Frequencies of the EM and CM modes at selected fields. (d) Relative intensities of the EM and CM modes normalized to their intensities at $H_g$, the horizontal dashed line represents $I_{EM} = I_{CM} = I_g$, which implies the coupling magnetic field.

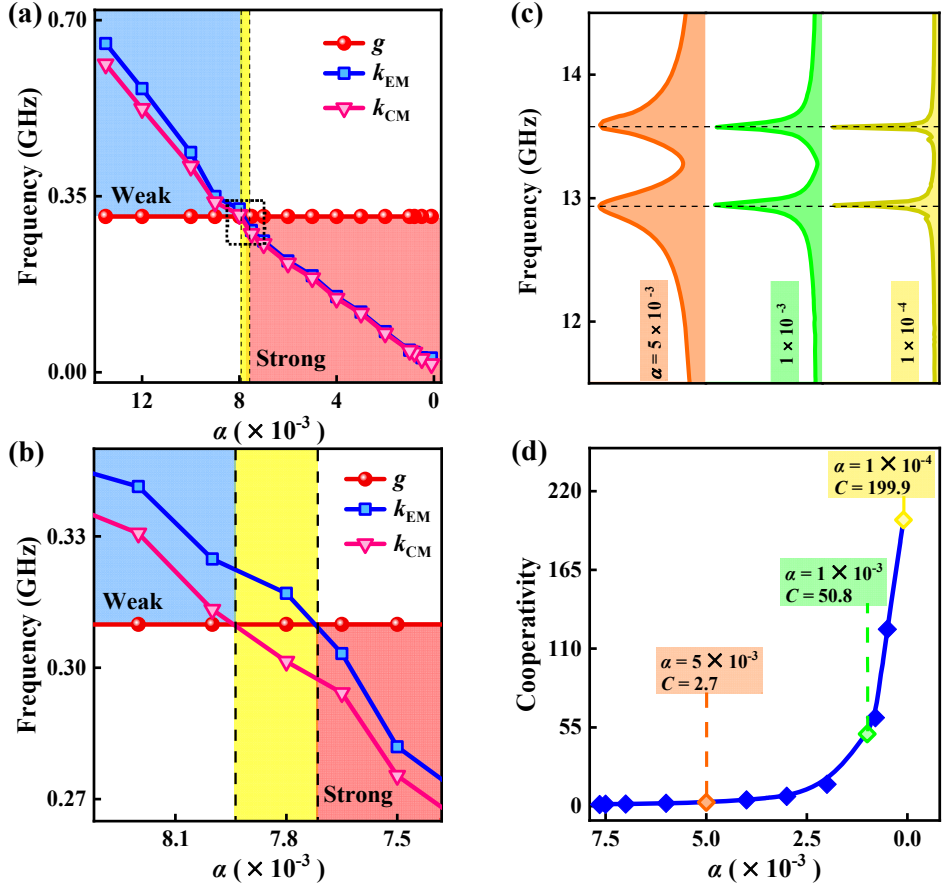

**Figure 3.** The coupling strength $g$, dissipation rates of EM and CM modes $k_{EM}$ and $k_{CM}$ in IHD ($l$ = 320 nm, $W$ = 240 nm, $\theta$ = 90°) as a function of the Gilbert damping parameter $\alpha$ at $H_g$ = 144 mT. The red, blue and yellow zones signify the system staying in the strong coupling, weak coupling and Purcell effect (or magnetically induced transparency) regimes at the corresponded damping values. (b) Zoomed-in view of damping-dependent coupling as labelled by a dotted square in (a). (c) Magnon spectra at coupling field $H_g$ = 144 mT for various damping values. (d) The magnon-magnon coupling cooperativity obtained from equation (1) at various damping values in the strong coupling regime.

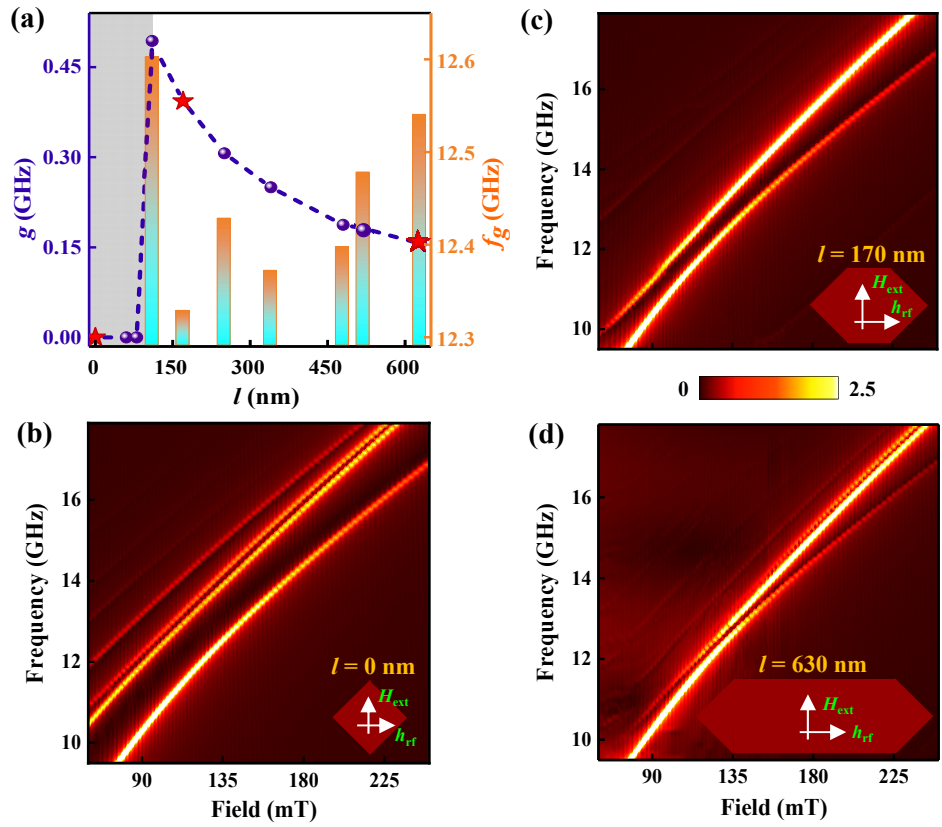

**Figure 4.** (a) Coupling strength $g$ (left axis, symbol + dashed line) and coupling center frequency $f_g$ (right axis, column charts) as a function of the length of IHDs with $W = 300$ nm and $\theta = 90°$. (b-d) Color plots of $f$-$H$ dispersion of magnons in IHDs with $l = 0$, 170, and 630 nm as labelled by stars in (a).

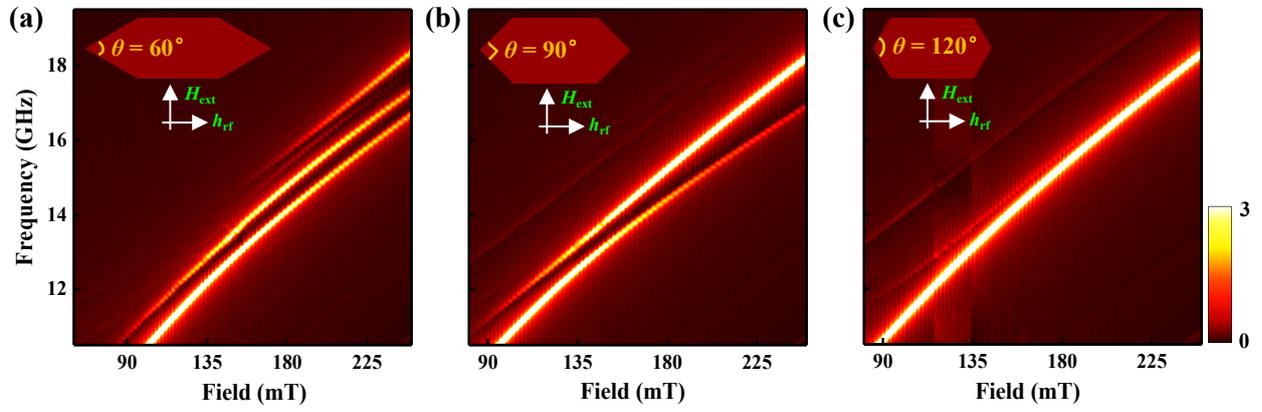

**Figure 5.** (a-c) Color plots of *f-H* dispersion of magnons in IHDs with *l* = 320 nm and *W* = 240 nm for $\theta$ = 60°, 90°, and 120°, respectively.